# THE DIFFRACTION MODEL AND ITS APPLICABILITY FOR WAKEFIELD CALCULATIONS

P. Hülsmann, W.F.O. Müller and H. Klein,
Institut für Angewandte Physik, Universität Frankfurt am Main, Germany

*Abstract*

The operation of a Free Electron Laser (FEL) in the ultraviolet or in the X-ray regime requires the acceleration of electron bunches with an rms length of 25 to 50 µm. The wakefields generated by these sub picosecond bunches extend into the frequency range well beyond the threshold for Cooper pair breakup (about 750 GHz) in superconducting niobium at 2 K. It is shown, that the superconducting cavities can indeed be operated with 25 µm bunches without suffering a breakdown of superconductivity (quench), however at the price of a reduced quality factor and an increased heat transfer to the superfluid helium bath.

This was first shown by wakefield calculations based on the diffraction model [1]. In the meantime a more conventional method of computing wake fields in the time domain by numerical methods was developed and used for the wakefield calculations [2]. Both methods lead to comparable results: the operation of TESLA with 25 µm bunches is possible but leads to an additional heat load due to the higher order modes (HOMs). Therefore HOM dampers for these high frequencies are under construction [3]. These dampers are located in the beam pipes between the 9-cell cavities. So it is of interest, if there are trapped modes in the cavity due to closed photon orbits.

In this paper we investigate the existence of trapped modes and the distribution of heat load over the surface of the TESLA cavity by numerical photon tracking.

## 1 INTRODUCTION

The use of a numerical photon tracking is justified by the fact, that we are only interested in photons above 750 GHz, which means a wavelength lower than 0.4 mm. From former investigations two important questions remain:

- Are there photons, which remain trapped within the cavity?
- Are there cavity surfaces with a pronounced local heat load?

The first point is of importance for the applicability of photon absorbers located in the beam pipes. The second point concerns the risk of quenching of cavities caused by concentrated photon absorption.

## 2 PHOTON TRACKING

For the calculations we used a numerical model of a mass point bouncing elastically through a two dimensional cut of the TESLA cavity like a classical billiard. The chaotic motion in these billiards is well investigated and used as model for eigenvalues of quantum systems. This is a suitable model for high frequency photons in a superconducting resonator, because of the very low surface resistance: In the frequency range below 750 GHz, which covers over 96% of the HOM energy of a 25 µm bunch, the surface resistance of niobium at 2 K is of some µΩ [2]. The characteristic number of hits $n$ depends on the impedance step at the surface:

$$\left(\frac{Z - Z_0}{Z + Z_0}\right)^n = \frac{1}{e} \quad (1)$$

So many thousands of reflections are possible. Above the Cooper pair threshold of 750 GHz the surface resistance rises immediately to 15 mΩ, but nevertheless some thousand reflections are possible.

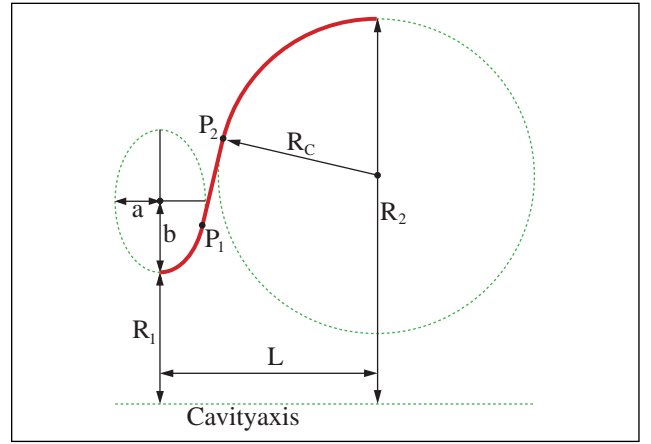

Figure 1: Elliptic and circular shape of TESLA cavity

Table 1: Geometric parameters of TESLA cavities

| geometric param. of half cell / [mm] | | inner cells | end cells (sym.) | asym. cavity end cell 1 | asym. cavity end cell 2 |
|---|---|---|---|---|---|
| length | L | 57.692 | 56 | 56 | 57 |
| iris radius | $R_1$ | 35 | 39 | 39 | 39 |
| cell radius | $R_2$ | 103.3 | 103.3 | 103.3 | 103.3 |
| elliptic radius 1 | a | 12 | 10 | 10 | 9 |
| elliptic radius 2 | b | 19.0 | 13.5 | 13.5 | 12.8 |
| circular radius | $R_C$ | 42.00 | 40.34 | 40.30 | 42.00 |
| total length | $L_T$ | | 1,035 | 1,036 | |
| beam pipe len. | $L_B$ | 346 | | | |

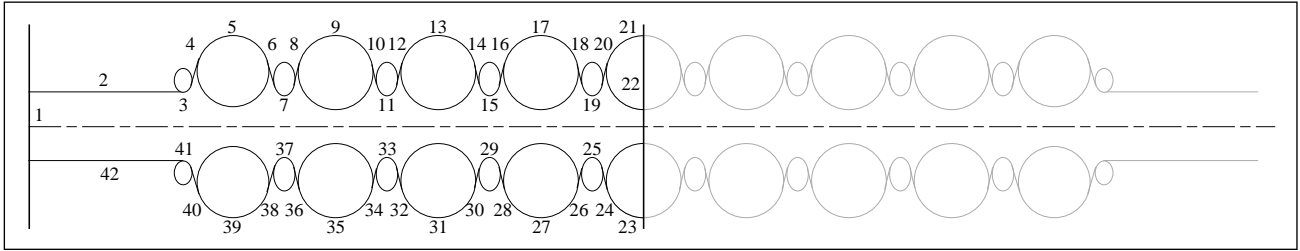

Figure 2: Definition of one half TESLA cavity with half beam pipe and symmetry planes (left and mid)

As can be seen in Fig. 1 the shape of the TESLA cavity is defined by a sequence of straight, round and elliptic graphical elements. The geometric data are given in table 1. Fig. 2 shows the geometry of one half 9-cell cavity consisting of 42 graphical elements. With a computer program the intersection of the photon track and each wall element is calculated and sorted with respect to the distance. Each intersection between photon trace and cavity contour is treated as a reflection process.

## 3 TRAPPED MODES

Following the concepts as described in [4], [5], [6] and [7] we assumed an entirely chaotic behavior for the elliptic shaped TESLA cavity. This was proven by numerical experiments with different starting conditions. For example, Fig. 3 shows two photon tracks (black and gray) started in the upper left corner of the cavity, both in parallel to the beam axis with slightly different offset. After a few reflections the tracks become completely different from each other. There is no need to distinguish between different classes of tracks, since photon tracks change between „whispering gallery orbits", „bouncing ball orbits" and orbits in parallel to the beam axis travelling from cavity to cavity through the beam pipe within some reflections.

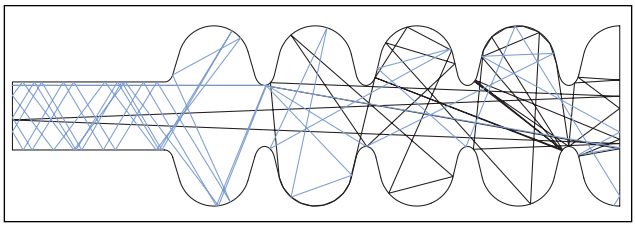

Figure 3: Two tracks with slightly different initial cond.

The next investigation is devoted to the problem of photons trapped in one cavity. Therefore in the shadow region of each cavity a photon source was established (see Fig. 4). This is the numerical analogon to the experiment with a light bulb in a cavity model made of copper [1].

The number of reflections until the escape through the beam pipe takes place is well below 100. Even from the middle cell (#5) photons leave the cavity within 64 reflections in the average (see Fig. 5). Towards the end cells this number is reduced to 23. Thus it is clearly shown, that there exists no location in the cavity, where photons are trapped.

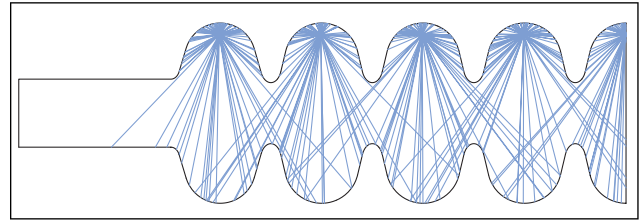

Figure 4: Photon sources in the shadow region of each cell in the center plane at r = 90 mm

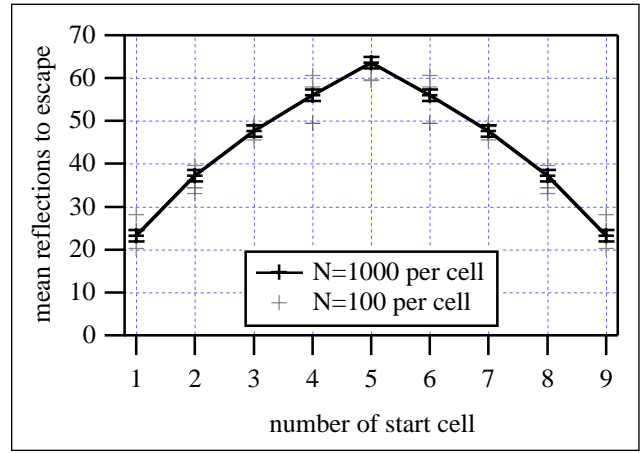

Figure 5: Reflections until escape through beam pipe

## 4 HEAT DISTRIBUTION

To study the process of heating a photon was injected into the half cavity. During the flight the photon experiences multiple reflections with the cavity wall. Fig. 6 shows the track of the photon after 1,000 reflections.

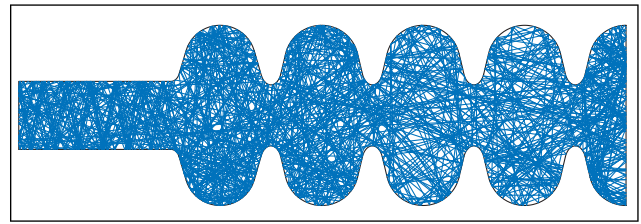

Figure 6: One photon track consisting of 1,000 reflections

As can be seen in Fig. 7 the large number of reflections and their chaotic structure have the beneficial effect, that the radiation power, which eventually has to be absorbed by the niobium, is distributed over the whole cavity surface. The oscillating characteristic in Fig. 7 is caused by the different shape of the reflecting objects: There are straight objects and objects with curvature alternating with each other, as can be seen in Fig. 1. And in the average the surface seen by the photon is smaller for bent objects. Despite this behavior the number of hits are equally distributed over the whole surface: There are about eight reflections per mm for a total number of 20,000 reflections.

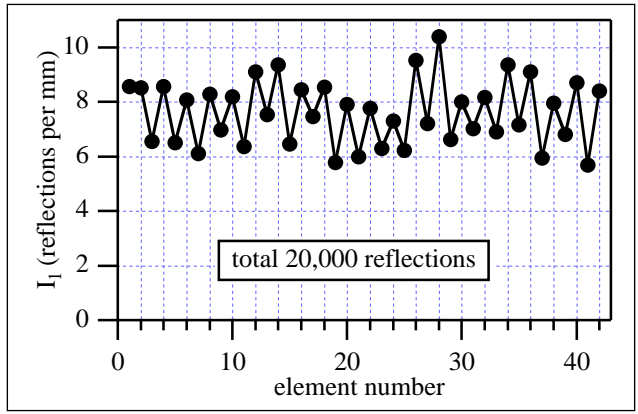

Figure 7: Hits per unit length versus geometric elements marked by numbers in Fig. 2

Due to the fact, that most of the diffracted radiation hits the first iris of the 9-cell cavity in a narrow ring shaped region close to the smallest diameter of the iris, a point like photon source at the iris edge was investigated. The initial condition with an emitting angle of $\pi/100$ is shown in Fig. 8:

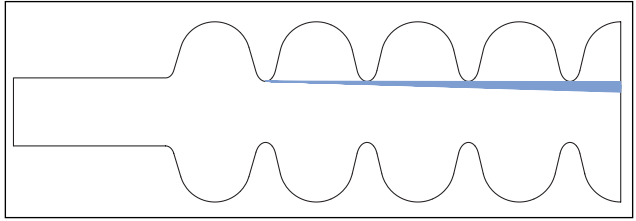

Figure 8: Point like photon emitter at the iris edge

The results in Fig. 9 show clearly, that the starting condition for multiple reflections is not of importance for the over all distribution. Only the symmetry planes #1 and #22 are hit slightly more often by the photons. This is caused by the fact, that both surfaces are hit in the very beginning of the process, where the motion is still deterministic. After five or ten reflections the motion is dominated by chaotic behavior.

## 5 CONCLUSION

The results of the numerical calculations have clearly shown, that the results delivered by a former light bulb experiment are in quite good agreement but a bit too pessimistic caused by the limited reflectivity of the copper model. Thus no problems caused by trapped photons are expected during FEL operation of the TESLA collider. The number of reflections needed to escape from a cavity is quite small compared to the number of reflections needed to absorb the photon in the wall. So the efficiency of high frequency photon absorbers for the THz region in the beam pipes can be taken for granted.

Furthermore it was shown, that the photon absorption is not concentrated on certain regions, e.g. the iris edges; the heat load is distributed over the whole cavity surface with a fluctuation of 25% only.

**(work supported by DESY / Hamburg and BMBF under contract 06 OF 841)**

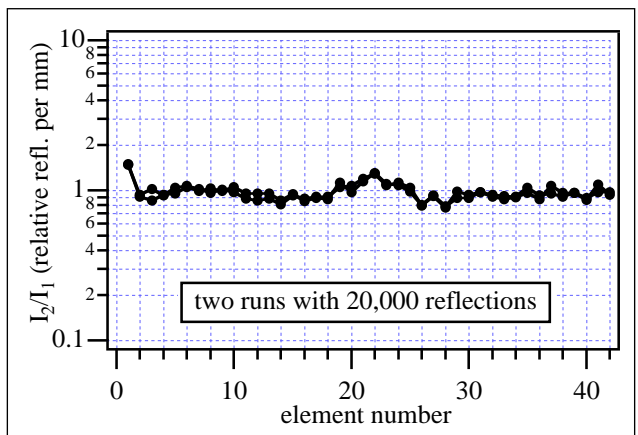

Figure 9: Comparison of hits per unit length for elements as marked in Fig. 2 by numbers